\documentclass[prd, aps, reprint, amsfonts, amssymb, amsmath, preprintnumbers, showpacs, nofootinbib]{revtex4-1}
\usepackage{graphicx, amsthm, multirow,soul}
\usepackage[citecolor=blue]{hyperref}
\usepackage[all]{hypcap}
\usepackage{url}
\usepackage{color}
\usepackage{subfigure}
\usepackage{float}
\usepackage{amsfonts}
\usepackage{bbold}

\newcommand{\equaref}[1]{Eq.~(\ref{#1})}

\newcommand{\figref}[1]{Fig.~\ref{#1}}

\newcommand{\subsecref}[1]{subsection~\ref{#1}}
\newcommand{\secref}[1]{Section~\ref{#1}}

\newcommand{\refref}[1]{Ref.~\cite{#1}}
\newcommand{\refsref}[2]{Refs.~\cite{#1}~and~\cite{#2}}

\newcommand{\deltasns}[1]{\ensuremath{\Delta(#1)}}

\begin{document}

\title{Predictions for Leptonic Mixing Angle Correlations and Non-trivial Dirac CP Violation from $A_5$ with Generalised CP Symmetry}

\author{Jessica Turner}
\email{jessica.turner@durham.ac.uk}

\affiliation{Institute for Particle Physics Phenomenology, Department of
Physics, Durham University, South Road, Durham DH1 3LE, United Kingdom}

\date{\today}

\begin{abstract}
Using discrete flavour group, $A_5$, combined with generalised CP, we study the mixing parameter correlations which arise from breaking to residual symmetries in the neutrino, $G_{\nu}=\mathbb{Z}_2\times \text{CP}$, and charged lepton sectors, $G_{e}=\mathbb{Z}_2$. By focusing on patterns that agree with current experimental data we demonstrate that non-trivial leptonic phases are predicted and discuss a number of  distinctive correlations between mixing parameters.

\end{abstract}

\preprint{IPPP/15/10, DCPT/15/20} 

\maketitle


\section{Introduction}

The  mixing structure of the three families  of the lepton sector has inspired many to use non-Abelian flavour symmetries to predict the   mixing angles and phases of the Pontecorvo-Maki-Nakagawa-Sakata (PMNS) matrix: three mixing angles $\theta_{13}$, $\theta_{12}$, $\theta_{23}$, the CP-violating  phase, $\delta$ and Majorana phases, $\alpha_{21}$ and $\alpha_{31}$. A significant experimental result in 2012 was the measurement of  $\theta_{13}$  \cite{An:2012eh,*An:2013uza,Ahn:2012nd,Abe:2011sj,*Abe:2011fz}.  This somewhat large value of $\theta_{13}$ ruled out many classes of flavour models that predicted small or zero $\theta_{13}$. These models lead to simple mixing patterns such as bi-maximal (BM) \cite{Fukugita:1998vn,Barger:1998ta,Davidson:1998bi,Altarelli:2009gn}, tri-bimaximal (TBM) \cite{Harrison:2002er,*Harrison:2003aw} or GR (golden ratio) \cite{Feruglio:2011qq}  which result from using small flavour groups such as $A_4$, $A_5$ and $S_4$ (an in-depth review of discrete groups can be found in  \refref{King:2013eh}). \\
In order to produce mixing patterns that accommodate experimental  data, the flavour model paradigm has shifted to include larger groups such as $\deltasns{96}$ \cite{Toorop:2011jn,Ding:2012xx}, $\deltasns{150}$ \cite{Lam:2012ga}, $\deltasns{600}$ \cite{Lam:2013ng} and $\deltasns{1536}$\cite{Holthausen:2012wt}. These non-Abelian discrete flavour groups cannot be a symmetry at the low-energy scale as leptonic masses are distinct. Therefore the flavour group must be broken into Abelian residual symmetries in the neutrino and charged lepton sectors. The structure of the Abelian residual symmetries is shaped by the larger non-Abelian flavour group and from these low-energy residual symmetries  leptonic observables can be predicted.  In general, there are two possible implementations of flavour symmetries and they are often referred to as  direct and semi-direct  (e.g see \refref{King:2013eh}). The distinction between the two approaches is the low energy residual symmetry of the Majorana mass matrix: in the direct approach the Klein group, $\mathbb{Z}_2\times \mathbb{Z}_2$, is  a subgroup of the underlying flavour symmetry whilst in the semi-direct approach,   $\mathbb{Z}_2$ emerges as a  residual symmetry of the flavour group. In the semi-direct models a continuous parameter is introduced, derived from the freedom to rotate in the degenerate subspace of the neutrino residual symmetry, allowing  the prediction of  a  non-zero $\theta_{13}$. There are several attractive features of implementing such an approach: firstly a UV-complete theory is not necessary in order to predict leptonic observables \cite{Hernandez:2012ra, Hernandez:2012sk,
Ballett:2013wya, Meloni:2013qda, Hanlon:2013ska, Petcov:2014laa,Ballett:2014dua}. Secondly, correlations between the observables  can be derived and provide specific signatures which allow the comparison of  a  range of models to experimental data  \cite{Antusch:2007rk,Ballett:2013wya,Meloni:2013qda,Hanlon:2013ska,Ballett:2014uia,
Petcov:2014laa, Girardi:2014faa, Ballett:2014dua}.\\
These models can  successfully  predict  mixing angles consistent with data and a Dirac phase, $\delta$. However, due to the constraints imposed on the mass matrices used to construct the PMNS matrix, a number of degrees of freedom cannot be eliminated and therefore these models cannot predict Majorana phases. By extending the flavour group to include  generalised CP (gCP) symmetry, the three mixing angles and three phases can be determined using a small number of input parameters \cite{Feruglio:2012cw}. This idea of combining CP with a flavour symmetry is not a recent one and was originally discussed in \cite{Harrison:2002kp,*Harrison:2002et,Grimus:2003yn,*Ferreira:2012ri,Farzan:2006vj} together with a  $\mu\mbox{-}\tau$ symmetry.  There have been a number of interesting works on the consistent relation between gCP and flavour symmetry \cite{Holthausen:2012dk,Feruglio:2012cw,Chen:2014tpa} and many plausible groups have been studied such as   $A_4$ \cite{Feruglio:2012cw,Ding:2013bpa}, $S_4$ \cite{Feruglio:2012cw,Feruglio:2013hia}, $\deltasns{96}$ \cite{Toorop:2011jn,Ding:2012xx}, $\deltasns{150}$ \cite{Lam:2012ga}, $\deltasns{600}$ \cite{Lam:2013ng}, $\deltasns{1536}$\cite{Holthausen:2012wt}, $\Delta(3n^2)$ \cite{Hagedorn:2014wha}, $\Delta(6n^2)$ \cite{Hagedorn:2014wha, Ding:2014ora} and most recently $A_5$ \cite{Li:2015jxa,DiIura:2015kfa,Ballett:2015wia}. In smaller groups such as $A_4$, $S_4$ and $A_5$ \cite{Feruglio:2012cw,Ding:2013bpa,Feruglio:2013hia,Li:2015jxa,DiIura:2015kfa,Ballett:2015wia}, it has been found that the leptonic phases are either trivial or maximal. Moreover, there are often share recurring patterns of predictions such as maximally CP violating  $\delta$ associated with maximal $\theta_{23}$, the origin of which was recently  discussed \cite{He:2015xha}. Applying the same framework with a  larger flavour group such as $\Delta(3n^2)$ or $\Delta(6n^2)$, leptonic phases are  non-trivially dependent upon the continuous parameter and can take values different from $0$, $\frac{\pi}{2}$, $\pi$ and $\frac{3\pi}{2}$. \\
The work presented in this paper is an extension of  the study in \cite{Ballett:2015wia} where a flavour group, $A_5$, combined with gCP is broken  into residual symmetries in the charged lepton, $G_{e}=\{\mathbb{Z}_3,\mathbb{Z}_5,\mathbb{Z}_2\times\mathbb{Z}_2\}$, and neutrino sectors, $G_{\nu}=\mathbb{Z}_2\times \text{CP}$. An additional motivation  to further explore the predictions of $A_5$ is, unlike other small groups such as $A_4$ and $S_4$, $A_5$ is anomaly safe \cite{Chen:2015aba}. In this work, we consider the possibility that  the same high energy symmetry breaks into low energy residual symmetries  $G_{e}=\mathbb{Z}_2$ and $G_{\nu}=\mathbb{Z}_2\times \text{CP}$.  By relaxing the possible combination of residual symmetries we find that non-trivial values of the leptonic phases can be accommodated and there are distinctive correlations between observables. \\
Throughout this work, we assume  our low energy effective theory is the Standard Model augmented by a Majorana mass term and we will use the following 3$\sigma$ global fit data \cite{Gonzalez-Garcia:2014bfa}

\begin{align*}
&7.85^{\circ}\leq\theta_{13}\leq9.10^{\circ},\quad
31.29^{\circ}\leq\theta_{12}\leq35.91^{\circ},\\
&38.2^{\circ}\leq\theta_{23}\leq53.3^{\circ}.
\end{align*}

The work presented in this paper is structured as follows: in \secref{sec:background} we present the assumptions of our theoretical framework; in \secref{sec:methodology} we discuss the construction of the PMNS matrix from  symmetry constraint and the derivation of our results. A number of representative predictions, along with an example, are given in \secref{sec:results} and finally we make concluding remarks in \secref{sec:conclusion}.

\section{Symmetries of the Model}\label{sec:background}
We  briefly review the theoretical framework of this study, where we have  closely followed the discussion of \cite{Ballett:2015wia}.  We first review the general concepts of flavour and generalised CP symmetry  and subsequently consider the consistent relations between these two symmetry transformations in preparation for constructing the PMNS matrix.

\subsection{Flavour Symmetry}\label{sec:flavour}
We assume there exists a  finite, discrete flavour symmetry, $G_{f}$,  at the high-energy scale. The purpose of this symmetry is to unify the three flavours of leptonic doublets into a single mathematical object: a three dimensional  irreducible representation of the flavour group, $\Psi$. The flavour group acts on  $\Psi$ such that 
\begin{equation}\Psi\to\rho(g)\Psi, \end{equation}
where $\rho(g)$ is a three-dimensional unitary representation of group element $g \in G_{f}$. The non-Abelian flavour symmetry must be broken at the  low-energy scale as leptonic masses are distinct. This implies that if a flavour symmetry is operational in the high-energy regime then only its Abelian residual symmetries would be observable at the scale of mass generation. Therefore, we assume that the non-Abelian flavour symmetry is broken into Abelian residual symmetries  in the charged lepton sector, $G_{e}$,  and  the neutrino sector, $G_{\nu}$. For  group elements $g_{e} \in G_{e}$ and $g_{\nu} \in G_{\nu}$,  the charged lepton and neutrino fields  transform under the residual symmetries according to
\begin{equation}\label{eq:trans1} e_\text{L} \to \rho\left(g_e\right) e_\text{L} \quad\text{and}\quad\nu_\text{L} \to \rho\left(g_\nu\right) \nu_\text{L}, \end{equation}
where generational indices have been suppressed. The transformations of \equaref{eq:trans1}  enforce constraints on the charged lepton and neutrino mass matrices,
\begin{align} \label{eq:ge}  \rho\left(g_e\right)^\dagger (m_e
m_e^\dagger)\rho\left(g_e\right)&=m_e m_e^\dagger,\\
 \rho\left(g_\nu\right)^\text{T}m_\nu
\rho\left(g_\nu\right)&= m_\nu. \label{eq:gnu}\end{align}
To deduce the possible forms of the residual symmetries, we must consider the largest  symmetry of each sector and  the structure  inherited from the larger non-Abelian flavour group. In the basis in which the charged lepton mass matrix is diagonal  and the masses are distinct, the  largest symmetry of this sector is $U(1)^{3}$. This is derived  from  the freedom to rephase the fields of each generation of the charged leptons. The most general discrete residual symmetry of this sector must be a subgroup of ${U(1)}^{3}$  and  is therefore a direct product of cyclic groups, $\mathbb{Z}_n$. The Abelian subgroups of $A_5$ that satisfy this condition are $\mathbb{Z}_5$, $\mathbb{Z}_3$, $\mathbb{Z}_2\times\mathbb{Z}_2$ and $\mathbb{Z}_2$ where the cases of $\mathbb{Z}_3,\mathbb{Z}_5$ and $\mathbb{Z}_2\times\mathbb{Z}_2$ have been studied in the analysis of \cite{Ballett:2015wia}. As we assume that neutrino are Majorana in nature, rather than Dirac type particles, their mass matrix is always invariant under a Klein symmetry,  $\mathbb{Z}_2\times\mathbb{Z}_2$.  Therefore the residual symmetry of the neutrino sector is the Klein group or a subgroup thereof.  

\subsection{ Generalised CP Symmetry}\label{sec:gCP}
In addition to the non-Abelian flavour symmetry operational at the high-energy scale, we assume there also exists a gCP symmetry. This  symmetry parity transforms and charge conjugates the field, as well as acting on its   generational indices \cite{Ecker:1987qp}. The gCP transformation acts on the multiplet of fields, $\Psi$, as
\begin{equation} 
\Psi\to X\Psi^{C}, 
\end{equation}
where $X$ is a unitary, symmetric matrix and $\Psi^{C}$ denotes the CP-conjugate of $\Psi$. We have chosen gCP  to be an involutory  meaning that two gCP transformations are equivalent to the identity: $XX^{*}=\mathbb{1}$.  If gCP remains a  symmetry of the charged lepton or neutrino sector, it must leave the mass terms invariant 
\begin{align} \label{eq:Xmnu} X^\text{T} m_\nu X &= m^*_\nu, \\
 X^\dagger
\label{eq:Xme}(m_e  m^\dagger_e) X &= (m_e
m^\dagger_e)^*. \end{align}
It has been demonstrated \cite{Branco:2011zb,Feruglio:2013hia} that  if gCP remains unbroken at the low-energy scale in both the charged lepton and neutrino sectors, then \equaref{eq:Xmnu} and \equaref{eq:Xme} are satisfied and consequently  no CP violating effects will be  observed. Therefore, in this paper we assume that gCP is  broken in the charged lepton sector and remains a preserved symmetry of the neutrino sector. \\
 In summary, the residual flavour and  gCP symmetries place a series of constraints on the charged lepton and neutrino mass matrices. These constraints shape the form of the diagonalising matrices , $U_{e}$ and $U_{\nu}$,  which in turn constrain the form of the PMNS matrix. The consistent interaction between the non-Abelian flavour group and gCP must be considered in order to determine which gCP transformations are physical. 

\subsection{Combining Flavour and gCP Symmetries}\label{sec:combine}
To illustrate the consistency between the  flavour and gCP symmetries, consider a multiplet of fields transforming under a gCP, flavour and subsequent gCP transformation
\begin{equation}\label{eq:autX}\Psi\to  X{\rho(g)}^{*}X^{*} \Psi  \equiv \rho(g^{\prime})\Psi.\end{equation}
In \cite{Holthausen:2012dk,Feruglio:2012cw}, the authors showed that in order to combine flavour and gCP symmetries consistently, gCP must act as an automorphism on the flavour group, $G_{f}$  (in \equaref{eq:autX}, gCP maps group element $g$ to another element $g^{\prime}$ such that the identity and group multiplication is respected). This idea was further developed by \cite{Chen:2014tpa}, which pointed out that for a physical CP transformation to occur, gCP should be an outer automorphism that maps the representations of the fields to their conjugate representations. Moreover, in a generic setting, these outer automorphisms of the group must be class-inverting automorphisms implying that \equaref{eq:autX} becomes    
\begin{equation}\label{eq:autX2}
 X {\rho(g)}^{*} X^{*} \Psi  \equiv \rho(h)\rho(g^{-1}) {\rho(h)}^{-1} \Psi,
\end{equation}
for $h \in G_{f}$. 
The detailed derivation of $X$ is discussed fully in \cite{Ballett:2015wia}, however in this work we will briefly summarise the group theoretic concepts that were considered. In order to find the forms of $X$ that constitute physical gCP transformations, the outer automorphism of the flavour group must be known. In general,  the outer automorphism group for $A_{n}$ where $n\leq5$, is $\mathbb{Z}_2$ (for more group theory insights see \refref{Rotman}). This implies there is only one non-trivial outer automorphism of $A_5$ and this maps elements of one conjugacy classes of order five to the other\footnote{$A_5$ contains 5 conjugacy classes: one for order one, two and three elements and two for order five elements.}. In addition to finding the non-trivial outer automorphism of the flavour group, $A_5$ has a special property that  simplifies the derivation of $X$. $A_5$ is an ambivalent group meaning each element is conjugate to its inverse. Applying this property of the group and choosing to work in a real representation, \equaref{eq:autX2} can be significantly simplified. Henceforth, a series of deductions can be made and it can be concluded that  
 the forms of $X$ that act on $A_5$ as class inverting, involutory automorphism are the Klein  group. 

\section{Methodology}\label{sec:methodology}
We first discuss the construction of the PMNS matrix from the symmetry constraints  and subsequently describe the method used to derive the correlations between observables.

\subsection{Constructing the PMNS matrix from symmetry considerations}\label{sec:construct}
The flavour and gCP symmetry constrain the form of the neutrino and charged lepton mass matrices. From these constraints the form of their diagonalising matrices, $U_{\nu}$ and $U_{e}$,  may be deduced and thus the PMNS matrix can be constructed: $U_{\text{PMNS}}=U_{e}^{\dagger}U_{\nu}$. Let us consider how to derive $U_{e}$ from the symmetry constraints. First, \equaref{eq:ge} can be re-expressed in the form of a  commutator: $[\rho(g_{e}),(m_e m_e^\dagger)]=0$. As the unitary representation $\rho(g_{e})$ commutes with the hermitian matrix $m_e m_e^\dagger$ there exists a unitary matrix, $U_{e}$,  that simultaneously diagonalises both. In the case that $\rho(g_{e})$ has degenerate eigenvalues, there is  not a unique diagonalising matrix of $\rho(g_{e})$ but rather an additional complex rotation can be performed in the degenerate subspace of $\rho(g_{e})$. Therefore, the most general form of the diagonalising matrix of $\rho(g_{e})$ and  $(m_e m_e^\dagger)$ is
\begin{equation} U_{e}=U_{l}R(\omega,\gamma), \end{equation}
where $U_l$ diagonalises $\rho(g_{e})$ and $R(\omega,\gamma)$ is an $\text{SU}(2)$ transformation in the degenerate eigenspace. It is worth stressing that  we allow for the existence of this complex rotation by  
 permitting  $G_{e}=\mathbb{Z}_2$.  This differs from the  analysis of \cite{Ballett:2015wia}  as their choice of $G_{e}$ had no such degenerate subspace. \\
In order to deduce the form of  $U_{\nu}$, we  consider  constraints from the flavour residual symmetry, gCP and the logical relation between the two symmetries.
The action of gCP on the neutrino residual symmetry, $\rho(g_{\nu})$,  maps these elements to their inverse:
\begin{equation}\label{eq:XZ2}X{\rho(g_{\nu})}^{*}X^{*}={\rho(g_{\nu})}^{-1}=\rho(g_{\nu}),\end{equation} 
where in the final step we have used the fact  $\mathbb{Z}_2$ elements are self inverse.  \equaref{eq:XZ2} can equivalently be viewed as  forming a direct product between $\mathbb{Z}_2$ and gCP. 
In \cite{Feruglio:2012cw} they showed that it is always possible to make a convenient basis change $\Omega$ ($X=\Omega\Omega^T$)  such that 
\begin{equation}\label{eq:Omega} (\Omega^{T}m_{\nu}\Omega)=(\Omega^{T}m_{\nu}\Omega)^{*}.\end{equation}
Therefore, this basis transformation ensures that $m_{\nu}$ is real valued. Moreover from \equaref{eq:gnu}, it can be seen that the diagonal from of  $\rho(g_{\nu})$ commutes with $(\Omega^{T}m_{\nu}\Omega)$, which implies that $m_{\nu}$ must be block diagonal.  To fully diagonalise the matrix of \equaref{eq:Omega}, an additional real rotation, $R(\theta)$, must be performed. From these  considerations,  $U_{\nu}$ may be written as 
\begin{equation} U_{\nu}=\Omega R(\theta).   \end{equation} 
 Using symmetry constraints  alone, the PMNS matrix may be written as
\begin{equation}\label{eq:PMNS}U_{\text{PMNS}}=R(\omega,\gamma){U_{l}}^{\dagger}\Omega R(\theta).\end{equation}

 For our chosen representation of $A_5$, $\Omega$ can take three possible forms
\begin{equation}
\begin{aligned}
\Omega_{12}&=\begin{pmatrix}
i&0&0\\
0&i&0\\
0&0&1
\end{pmatrix},\quad
&\Omega_{13}&=\begin{pmatrix}
i&0&0\\
0&1&0\\
0&0&i
\end{pmatrix}\\\quad\text{and}\quad
\Omega_{23}&=\begin{pmatrix}
1&0&0\\
0&i&0\\
0&0&i
\end{pmatrix},
\end{aligned}
\end{equation}
which have been fully derived in \cite{Ballett:2015wia}. 

\begin{figure*}[t!]
\label{fig:exampleA}
  \centering
    \includegraphics[width=0.3\textwidth]{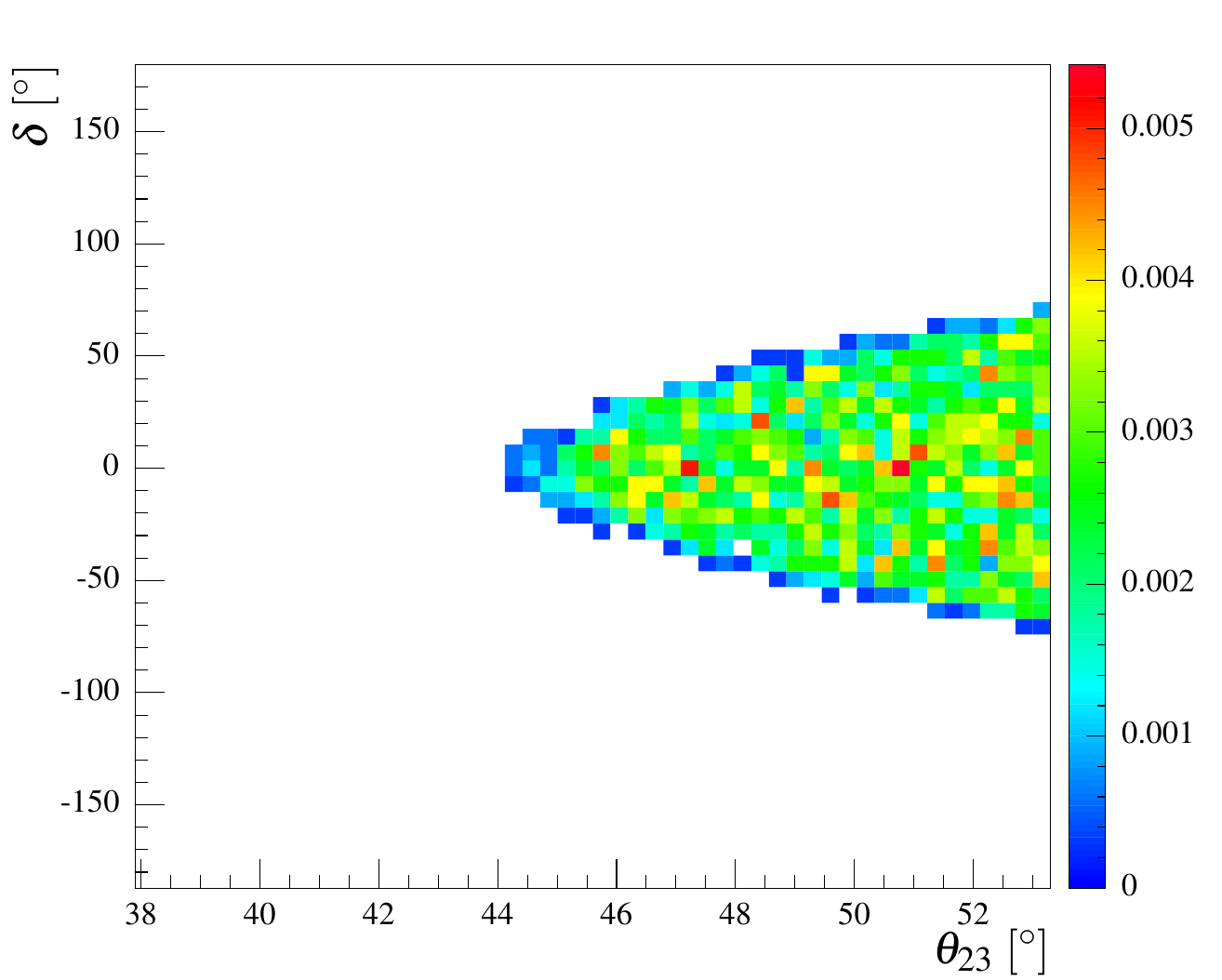}
 \includegraphics[width=0.3\textwidth]{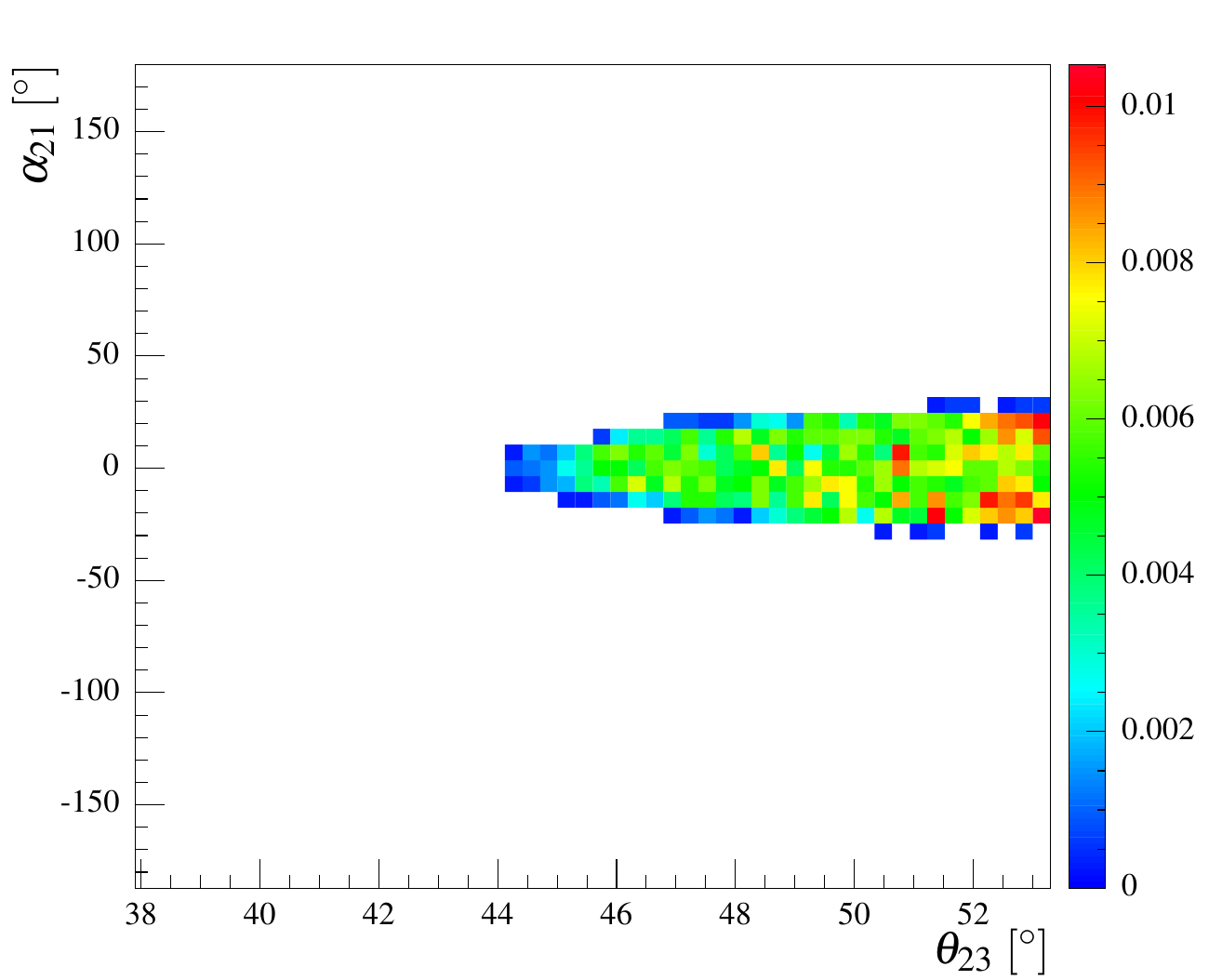}
 \includegraphics[width=0.3\textwidth]{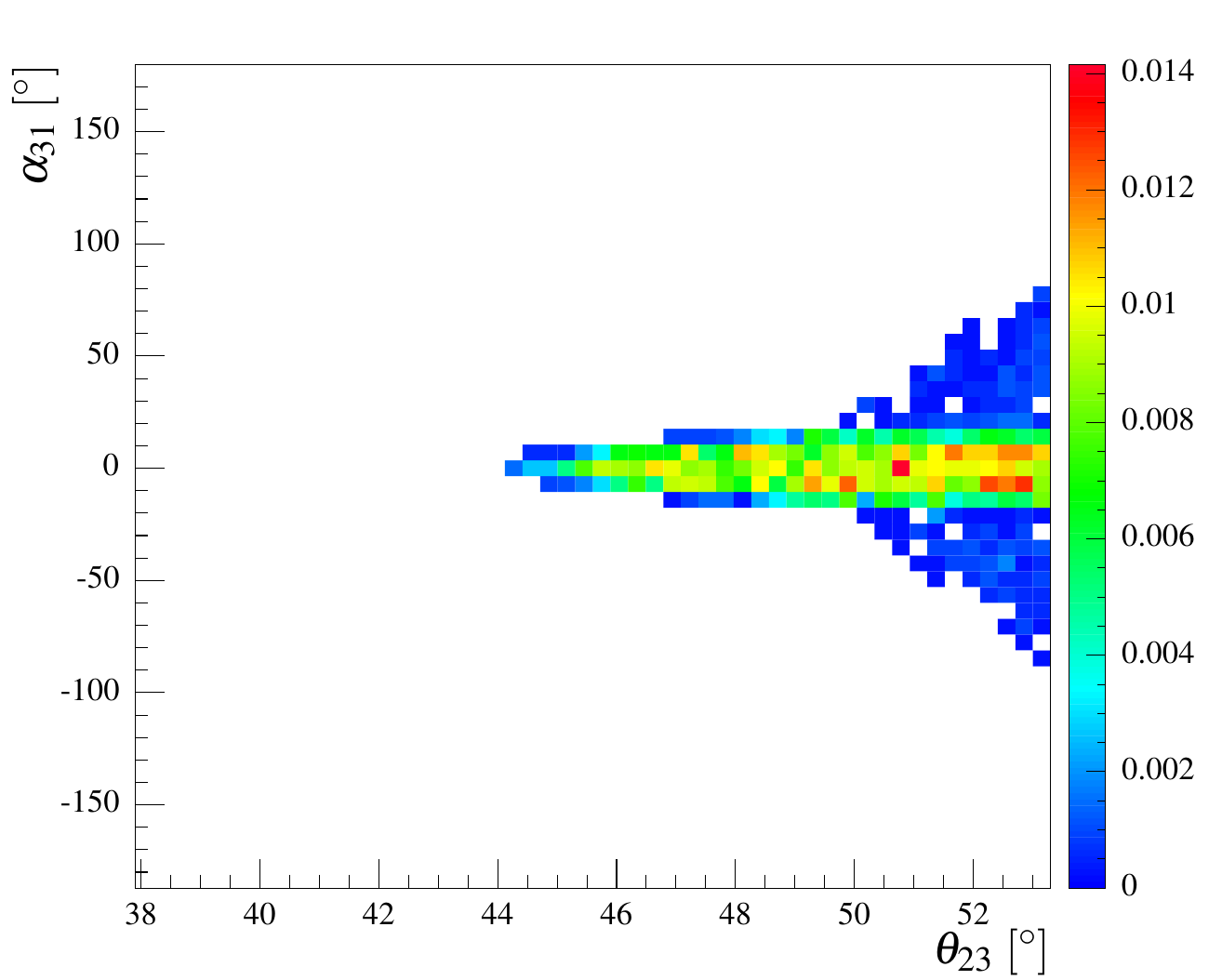}
 \includegraphics[width=0.3\textwidth]{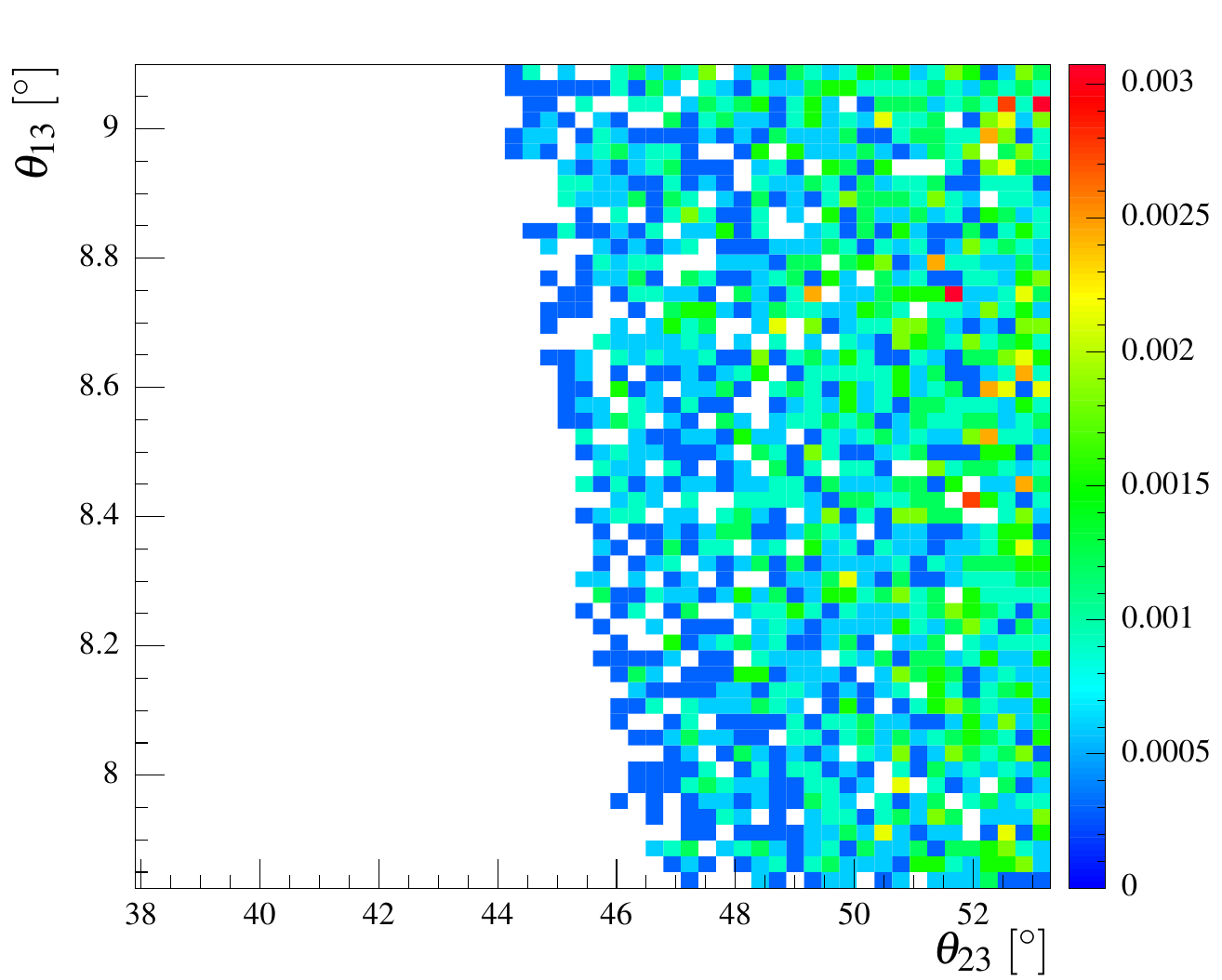}
 \includegraphics[width=0.3\textwidth]{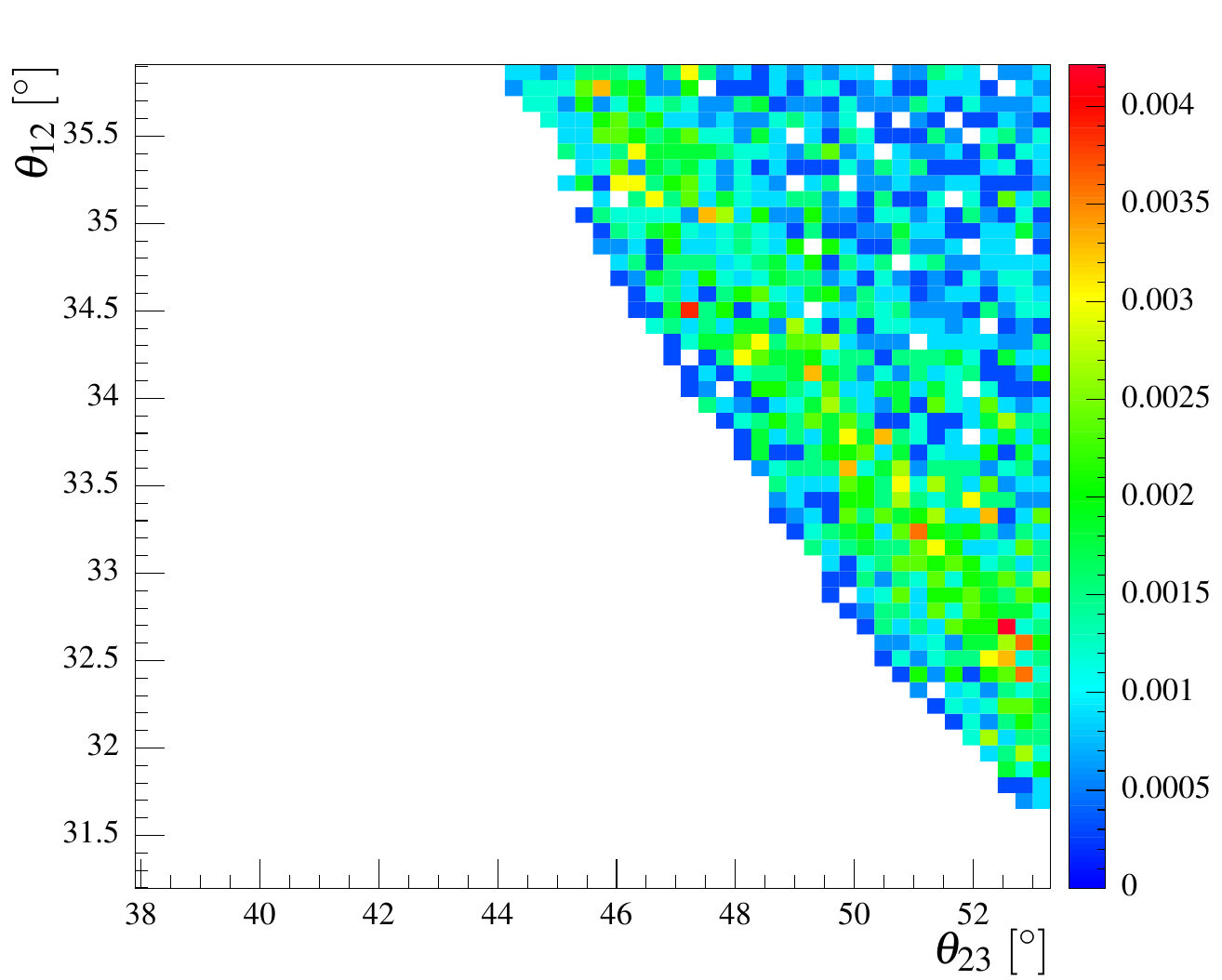}

  \caption{Two-dimensional histograms showing the phases  and mixing angles as a function of $\theta_{23}$ for  predictions that agree to a 3$\sigma$ level with global fit data \cite{Gonzalez-Garcia:2014bfa}.  The colour chart shows relative frequencies of solutions where red (dark blue) represents a higher (lower)  frequency.}
\end{figure*}

\subsection{Derivation of predictions for leptonic mixing parameters}\label{sec:method}

In the work of \cite{Ballett:2015wia},  the combination of Abelian residual symmetries studied were  $G_{e}=\{\mathbb{Z}_3$, $\mathbb{Z}_5$, $\mathbb{Z}_2\times\mathbb{Z}_2\}$ and  $G_{\nu}=\mathbb{Z}_2\times \text{CP}$. The observables are a function of one continuous parameter,  derived from the freedom to make a real rotation in the  degenerate subspace of the $\mathbb{Z}_2$ residual symmetry in the neutrino sector. Moreover, the authors of  \cite{Ballett:2015wia}, considered the combination of $G_{\nu}=\mathbb{Z}_2\times\mathbb{Z}_2\times \text{CP}$ and $G_{e}=\mathbb{Z}_2$. In this case, each observable would be a function of two input parameters which is obtained from the ability to make an $\text{SU}(2)$ transformation in the degenerate eigenspace of the $\mathbb{Z}_2$ charged lepton residual symmetry. This was fully explored and there were no predictions that agreed with data at the  3$\sigma$ level.  A further consideration of $G_{e}=\mathbb{Z}_2$ and  $G_{\nu}=\mathbb{Z}_2\times \text{CP}$ was proposed in \cite{Ballett:2015wia}. Moreover, predictions that agreed to a 3$\sigma$ level with global fit data were found, however these correlations were not analysed further. \\
In this work, we aim to develop on this combination of residual symmetries that was originally described in \cite{Ballett:2015wia} and explore  the non-trivial correlations between leptonic observables. This consideration allows for two additional continuous parameters and therefore each observable is a function of three input parameters: $\theta$, $\omega$ and $\gamma$. \\
For each of the fifteen  charged lepton residual symmetries, $\rho(g_{e})$, we  construct a diagonalising matrix: $U_{e}=U_{l}R(\omega,\gamma)$ where we have fixed the degenerate subspace, $R(\omega,\gamma)$. To construct $U_{\nu}$ we have explored all possible combinations of $\Omega R(\theta)$ and from this, we can construct $U_{\text{PMNS}}$. For each constructed PMNS matrix, we considered the arbitrary ordering of the diagonalising matrices and therefore appropriately permuted the rows and columns. For each PMNS matrix, the three continuous parameters ($\theta$, $\omega$,$\gamma$) are randomly scanned over the range $[0,\pi]$ and the three mixing angles are calculated for each point in the phase space.  Subsequently, only points that simultaneously agree to a  3$\sigma$ level with global fit data \cite{Gonzalez-Garcia:2014bfa} are retained and from these points the phases are calculated. We have chosen to present our results in terms of  correlations between the leptonic phases, $\theta_{13}$, $\theta_{12}$ as a function of $\theta_{23}$.

\section{Results}\label{sec:results}
There  is a varied  programme of currently running and planned experiments that aim to increase  precision in the measurement of   a number of the oscillation parameters. In the near term, accelerator long-baseline experiments such as T2K \cite{Abe:2014tzr} and NO$\nu$A \cite{Patterson:2012zs} aim at improving  the current measurements of  parameters such as  $\theta_{23}$, $\delta$ and $\Delta {m_{32}}^{2}$. 
In the longer term,  accelerator facilities such as DUNE \cite{Johnson:sub1}  and T2HK \cite{Abe:2014oxa} hope to further increase the sensitivity to these oscillation parameters.  T2HK will have the  ability to resolve $\delta$ to a 1$\sigma$ uncertainty of $19^{\circ}$ for all allowed values (using an integrated beam power of $7.5$MW seconds of exposure with $1.56\times 10^{22}$ protons on target).  Moreover, using a 10kt detector and  expected knowledge from T2K and NO$\nu$A, would allow DUNE to achieve a 3$\sigma$ sensitivity for  detecting CP-violation in $50\%$ of $\delta$ values.\\
In conjunction, future medium baseline reactor experiments such as RENO-50 \cite{Kim:2014rfa} and JUNO \cite{Li:2014qca}, aim to better the  measurement of $\theta_{12}$. These experiments utilise the survival probability of  electron anti-neutrinos, which are copiously produced in fission reactors, to determine the mass ordering and make sub-percent measurements of $\theta_{12}$. \\
The  determination of the nature of the neutrino remains of fundamental importance and neutrinoless double beta decay ($\nu0\beta\beta$) experiments   such as GERDA, CUORICINO, EXO-200 and KamLAND-Zen hope to explore CP-conserving upper boundary of the inverted ordering region.  The decay rate of this rare process  is proportional to the effective Majorana mass, $m_{ee}$ (see e.g \cite{Petcov:1993rk,Pascoli:2003ke,Vissani:1999tu,Pascoli:2005zb,Choubey:2005rq,Simkovic:2010ka,Dell'Oro:2014yca}) and the  values of $m_{ee}$ are influenced by the combinations of phases, $e^{i\alpha_{21}}$ and $e^{i\left( \alpha_{31}-2\delta   \right)}$. We will comment on some specific predictions which have particularly  relevant consequences  for  $\nu0\beta\beta$. However, due  to the ambitious plans to improve measurement of $\delta$, $\theta_{12}$ and $\theta_{23}$ by a range of complementary  neutrino oscillation experiments, we will mainly focus  on the mixing angle and  $\delta\mbox{-}\theta_{23}$ correlations.\\
Using the symmetry construction of \secref{sec:background}, each PMNS matrix  is a function of three continuous parameters and therefore we find a large number of cases that agree to a 3$\sigma$ level with global fit data and we have in the order of fifty different predictions. For illustrative purposes we provide an explicit example of one such prediction   in \subsecref{sec:exampleA}.  We group the remainder of our selected predictions into categories according to the octant of the $\theta_{23}$: the  lower octant cases are discussed in  \subsecref{sec:LOP}, upper octant  in \subsecref{sec:UOP}  and finally cases that span both octants are discussed in  \subsecref{sec:BOP}.
%

\subsection{ An Example } \label{sec:exampleA}
 We utilise the group   representations of \refsref{Everett:2008et}{Ding:2011cm}. To construct $U_{e}$, let us first consider a $\mathbb{Z}_2$ group elements of $A_5$ in the three-dimensional real representation
\begin{equation}\label{eq:Z2}
\mathbb{Z}_2=\frac{1}{2}\begin{pmatrix}
-1&\phi&\frac{-1}{\phi}\\
\phi&\frac{1}{\phi}&-1\\
\frac{-1}{\phi}&-1&-\phi
\end{pmatrix},
\end{equation}
where $\phi=\frac{\left(1+\sqrt{5}\right)}{2}$ is the golden ratio. A diagonalising matrix of \equaref{eq:Z2} is 
\begin{equation}
U_{l}\sim\begin{pmatrix}
 0.665 & -0.555& -\frac{1}{2} \\
 -0.58 & -0.025 & -\frac{\phi}{2} \\
 0.461 & 0.832& -\frac{1}{2\phi} \\
\end{pmatrix},
\end{equation}
 where the degenerate eigenvalues of the matrix of \equaref{eq:Z2} are in the 12-plane. Therefore, $U^{\dagger}_{e}$ takes the form
\begin{equation}\label{eq:Ue}
\resizebox{\linewidth}{!}{
$
U^{\dagger}_{e}\sim\begin{pmatrix}
c_{\omega} & e^{i \gamma}s_{\omega} & 0 \\
 -e^{-i \gamma } s_{\omega} &c_{\omega} & 0 \\
 0 & 0 & 1 \\
\end{pmatrix}
\begin{pmatrix}
0.665 & -0.58 &  0.461 \\
 -0.555& -0.025 & 0.832 \\
 -\frac{1}{2} & -\frac{\phi}{2}& -\frac{1}{2\phi} \\
\end{pmatrix},
$
}
\end{equation}
where $c_{\omega}\equiv\cos\left(\omega\right)$ and $s_{\omega}\equiv\sin\left(\omega\right)$. Not all combinations of $\Omega$ and $R(\theta)$  produce predictions  within  3$\sigma$ of the global fit data, however,
 one such combination that does is

\begin{equation}\label{eq:Unu}
U_{\nu}=\begin{pmatrix}
1&0&0\\
0&i&0\\
0&0&i
\end{pmatrix}
\begin{pmatrix}
 1 & 0 & 0 \\
 0 &c_{\theta}& s_{\theta}\\
 0 & -s_{\theta}&c_{\theta}\\
\end{pmatrix}.
\end{equation}

Combining \equaref{eq:Ue} and \equaref{eq:Unu}, we construct the PMNS matrix and perform a random scan over
the three continuous parameters ($\theta, \omega,\gamma$)  in the interval $[0,\pi]$. The points of this parameter space that agree to a 3$\sigma$ level with  global fit data are retained and the phases  are calculated. We  use the Particle Data Group parametrisation  to obtain the mixing angles and phases \cite{Agashe:2014kda}. In  \figref{fig:exampleA}, we plot the three leptonic phases, $\theta_{12}$ and $\theta_{13}$ as a function of $\theta_{23}$. This PMNS matrix requires  $44.2^{\circ}\leq \theta_{23}\leq 53.3^{\circ}$ and the $\delta$ phase reaches a maximal value of $69^{\circ}$ for large values of $\theta_{23}$ ($53^{\circ}$). A CP conserving value of $\delta$ is possible for all viable $\theta_{23}$ and  for maximal $\theta_{23}$, $-20^{\circ}\leq\delta\leq20^{\circ}$. From \figref{fig:exampleA}, there appears to be no preferred $\delta$ phase within the viable parameter space.  The values of the Majorana phases range from $\alpha_{21} \sim\pm 25^{\circ}$ and $\alpha_{31} \sim\pm 80^{\circ}$. In the case of $\alpha_{31}$, small values ($<15^{\circ}$) are strongly preferred over the whole range $\theta_{23}$. However for $\theta_{23}>49^{\circ}$, $\alpha_{31}$ can take large values. The consequences of this prediction on neutrinoless double beta  ($\nu0\beta\beta$) decay would be interesting to explore. The magnitudes of  $\alpha_{21}$ and $\left(\alpha_{31}-2\delta\right)$  of this prediction can be small and this  results in  little cancellation between the mass terms of $m_{ee}$. This would imply the prediction for $m_{ee}$  can be close to the CP-conserving upper boundary of the inverted ordering region, which experiments  hope to explore. Therefore, it would be feasible to use $\nu0\beta\beta$ decay to study this particular prediction.  \\
As the 3$\sigma$ range of $\theta_{13}$ is highly constrained compared with the other mixing angles, there is little discernible structure in the $\theta_{13}\mbox{-}\theta_{23}$ correlation, however   the $\theta_{12}\mbox{-}\theta_{23}$ dependence has greater predictivity. For near maximal values of $\theta_{23}$, $\theta_{12}$ is predicted to be at the very upper boundary of its 3$\sigma$ range ($\sim 36^{\circ}$). For larger values of $\theta_{23}$, close to the upper 3$\sigma$ boundary, the range of predicted $\theta_{12}$ increases ($31.6^{\circ}\mbox{-}35.9^{\circ}$).   Although, for most viable values of $\theta_{23}$ there are a range of $\theta_{12}$ predictions, the density of solutions clusters near the boundary of the viable region of the $\delta\mbox{-}\theta_{23}$ parameter space.

\subsection{Lower Octant Predictions}\label{sec:LOP}

\begin{figure*}[h!]\label{fig:LOP}
\centering
  \includegraphics[width=1.0\textwidth]{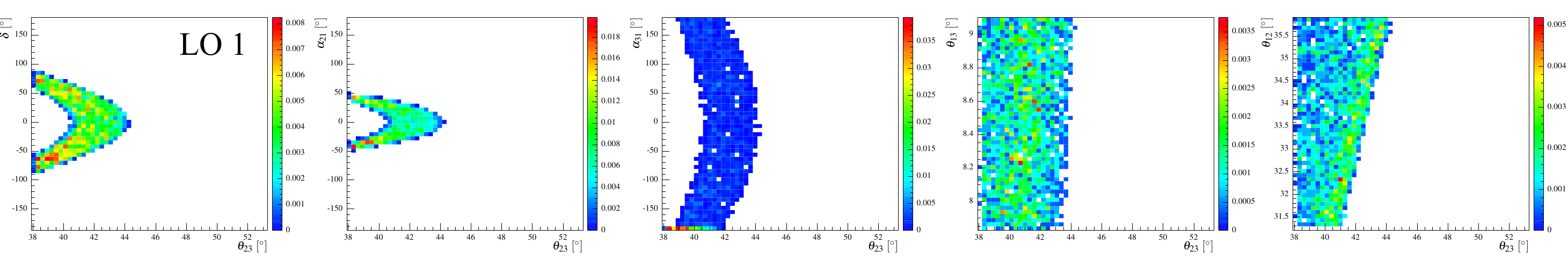}
  \includegraphics[width=1.0\textwidth]{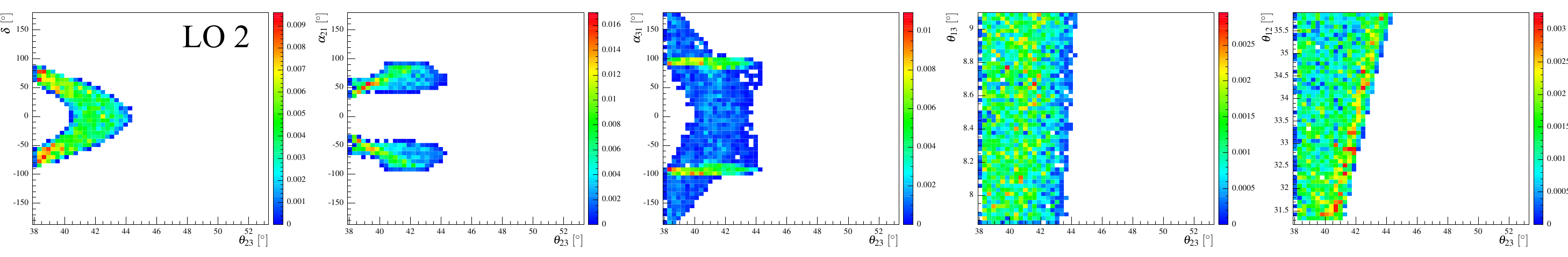}
  \includegraphics[width=1.0\textwidth]{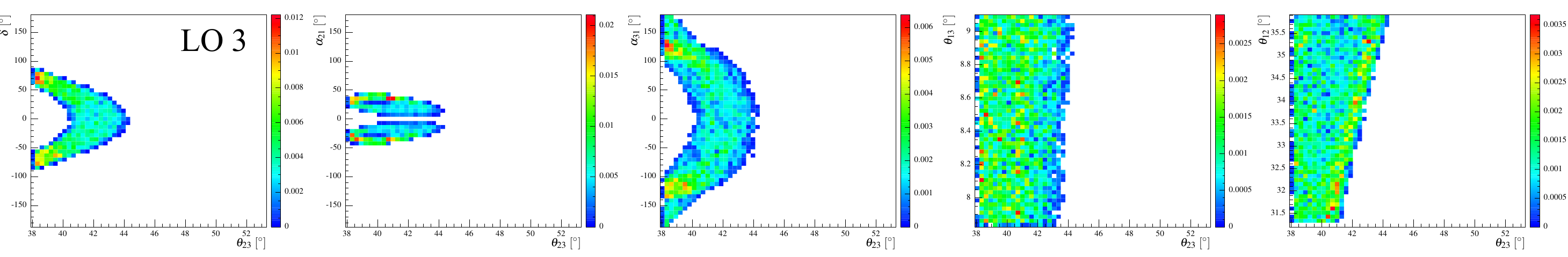}
  \includegraphics[width=1.0\textwidth]{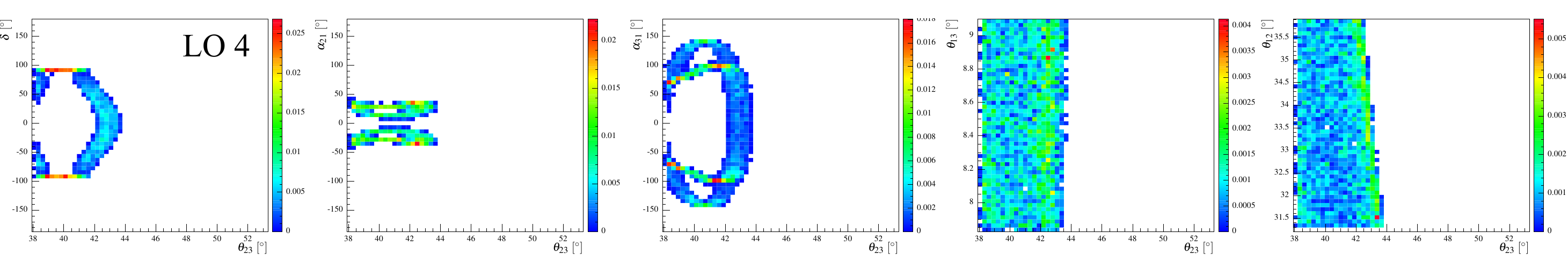}
  \includegraphics[width=1.0\textwidth]{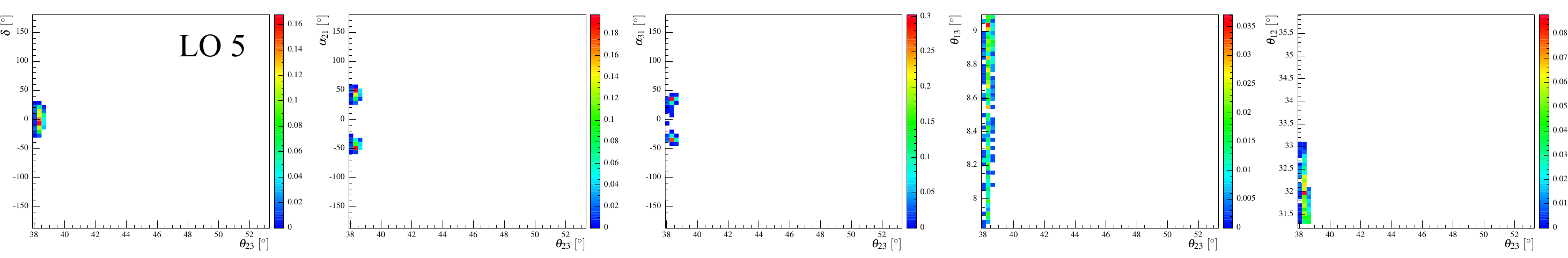}
\caption{Two-dimensional histograms showing the phases  and mixing angles as a function of $\theta_{23}$ for  predictions that  agree to a 3$\sigma$ level with global fit data\cite{Gonzalez-Garcia:2014bfa}  The colour chart shows relative frequencies of solutions where red (dark blue) represents a higher (lower)  frequency. Each prediction is labelled 'lower octant'  (LO) 1-5.}
\end{figure*}

The chosen lower octant predictions are presented in \figref{fig:LOP} and it can be seen that the  possible range of $\theta_{23}$ values differs between the various cases:  LO 1-3 have viable predictions for the entire lower octant  ($38.8^{\circ}\mbox{-}45^{\circ}$) whilst LO 4 is somewhat more constrained as $\theta_{23}$ spans only $3^{\circ}$ ($40^{\circ}\mbox{-}43^{\circ}$).  LO 5 is the most highly constrained and therefore most easily testable with $38^{\circ}\leq\theta_{23}\leq38.8^{\circ}$.  LO 1-3 share the same $\delta\mbox{-}\theta_{23}$ correlation, which  attains a maximal  $\delta$ ($85^{\circ}$)  for  $\theta_{23}$ close to the lower 3$\sigma$ allowed region. The CP conserving values of $\delta$ requires $40.6^{\circ}\leq\theta_{23}\leq44.2^{\circ}$. In the case of LO4, although the $\delta\mbox{-}\theta_{23}$ correlation structure  is similar to that of LO 1-3, the maximal value of $\delta$  is slightly greater reaching $90^{\circ}$ and solutions tend to cluster at these points. In contrast to LO 1-4, the $\delta$ value of LO 5 is close to zero however it reaches a maximum of $26^{\circ}$. \\
LO 1-3 have a common  $\theta_{12}\mbox{-}\theta_{23}$ dependence: for values of $\theta_{23}$ close to the lower 3$\sigma$ boundary, all values of $\theta_{12}$ are allowed. For near maximal   $\theta_{23}$, the $\theta_{12}$ prediction becomes increasing constrained: for example for the current best fit value of $\theta_{23}$ ($42.3^{\circ}$) \cite{Gonzalez-Garcia:2014bfa}, only values of $33.5^{\circ}\leq\theta_{12}\leq35.91^{\circ}$ are predicted. 
In the case of LO 4, for $38.8^{\circ}\leq\theta_{23}\leq42.3^{\circ}$, the predicted  $\theta_{12}$ spans the 3$\sigma$ range of $\theta_{12}$. Similarly to LO 1-3, the range of predicted  $\theta_{12}$ becomes more constrained for near maximal $\theta_{23}$ (smaller $\theta_{12}$ is preferred).  Using the $\theta_{12}\mbox{-}\theta_{23}$ correlation as a means of differentiating between LO 1-3 and LO 4 would be problematic in the regions $\theta_{23}\leq41^{\circ}$, as the predictions are indistinguishable. As the viable parameter space of LO 5 is significantly smaller than that of the previous four predictions, there is no discernible correlation between  $\theta_{12}$ and $\theta_{23}$; in this regards its most discriminating feature is that $\theta_{12}$ can only range between $31^{\circ}\mbox{-}33^{\circ}$.\\
The Majorana phases are the only observables that differs amongst LO 1-3. It is worth noting that LO 1 is the only lower octant prediction of this sample that has a CP conserving value of $\alpha_{21}$. It would be an interesting future study to investigate the effect  that this would have on $\nu0\beta\beta$ decay and   feasibility of discriminating between predictions. \\
In summary, there are several general features which are shared amongst these cases; the most striking of these is the prediction of  non-trivial leptonic phases. Moreover,  the $\delta$ and $\alpha_{21}$ phases are bound between $\pm90^{\circ}$. If $\delta$ is measured to be maximally CP violating, as hinted  at by T2K \cite{Abe:2015awa}, the only remaining viable prediction is LO 4. Some predictions cannot be discriminated between by using $\delta$ and $\theta_{12}$ alone and  access to the Majorana phases is necessary. Moreover, the ability to discriminate between LO 4 and LO 1-3 is highly dependent upon the value of $\theta_{23}$:  in the scenario of maximal or near maximal  $\theta_{23}$,  this is possible. Of the cases presented, LO 5 is the most easily testable as its $\theta_{23}$ values are highly constrained and lie at the extreme lower boundary of the 3$\sigma$ range.

\subsection{Upper Octant Predictions}\label{sec:UOP}

\begin{figure*}[h!]\label{fig:UOP}
\centering
  \includegraphics[width=1.0\textwidth]{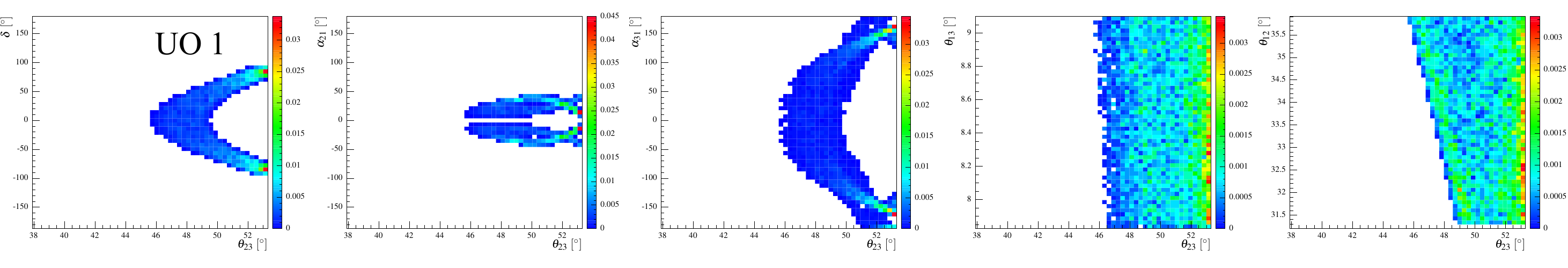}
 \includegraphics[width=1.0\textwidth]{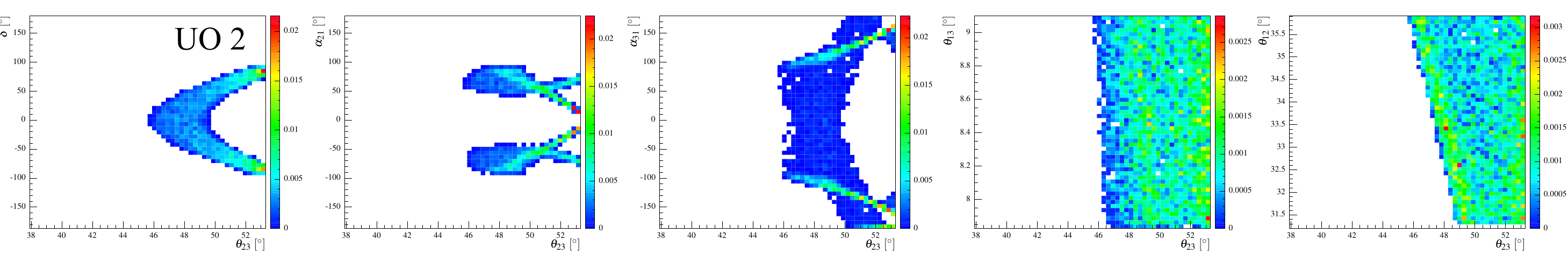}
  \includegraphics[width=1.0\textwidth]{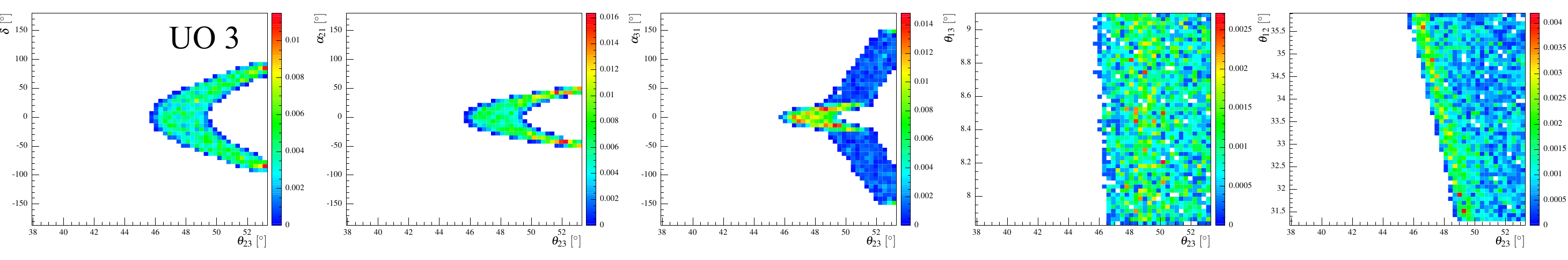}
  \includegraphics[width=1.0\textwidth]{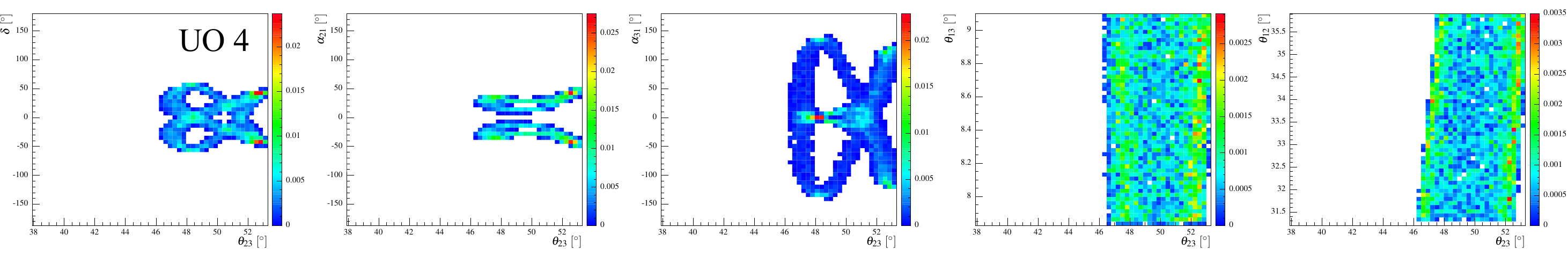}
  \includegraphics[width=1.0\textwidth]{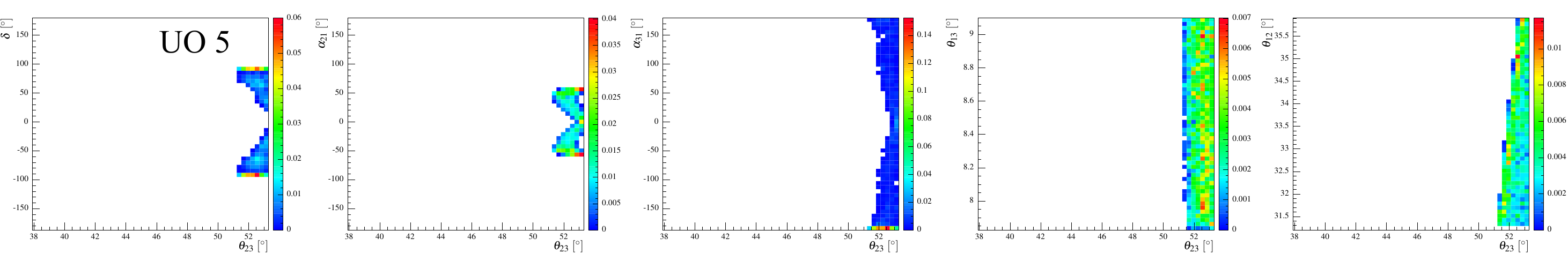}

\caption{Two-dimensional histograms showing the phases  and mixing angles as a function of $\theta_{23}$ for  predictions that  agree to a 3$\sigma$ level with global fit data\cite{Gonzalez-Garcia:2014bfa}  The colour chart shows relative frequencies of solutions where red (dark blue) represents a higher (lower)  frequency. Each prediction is labelled 'upper octant'  (UO) 1-5.}
\end{figure*}

Similarly to the lower octant results, we have chosen three cases (UO -1-3) presented in \figref{fig:UOP},  for which the mixing angle and $\delta$ phase correlations are indistinguishable and  only the Majorana phases differ. UO 1-3 share the feature of  viable predictions over the entire upper octant ($45^{\circ}\mbox{-}53.3^{\circ}$).
The $\theta_{23}$ allowed range UO 4 is slightly more constrained with $46.3^{\circ}\leq\theta_{23}\leq53.3^{\circ}$. UO 5 is an analogous case to LO 5, where its $\theta_{23}$ prediction span is small and occurs at the very upper limit of the 3$\sigma$ boundary, $51.2^{\circ}\leq\theta_{23}\leq53.3^{\circ}$.\\
Maximal CP violation is possible in UO 1-3 and UO 5, however the $\delta\mbox{-}\theta_{23}$ correlations structure differs between cases. UO 1-3 share the same pattern where the maximal $\delta$ value ($90^{\circ}$) occurs for large $\theta_{23}$ values and CP conserving values of $\delta$ are associated with $45.6^{\circ}\leq\theta_{23}\leq48.4^{\circ}$. The $\delta$ correlation of UO 5 differs significantly from UO 1-3 as 
 CP conserving values of $\delta$ are not predicted and maximal $\delta$ favoured.  In the case of UO 4, the   correlation structures are particularly distinctive and unlike the previously discussed cases, the maximal $\delta$ value ($55^{\circ}$) is much smaller. A unique aspect of LO 4 is that there are two distinct regions of $\theta_{23}$ where CP conserving values of $\delta$ can occur: $47.4^{\circ}\leq\theta_{23}\leq49.2^{\circ}$ and $51^{\circ}\leq\theta_{23}\leq52^{\circ}$. \\
In regards to the  $\theta_{12}\mbox{-}\theta_{23}$ correlation of UO 1-3, all regions of the 3$\sigma$ range of $\theta_{12}$ are allowed for $49^{\circ}\leq\theta_{23}\leq53.3^{\circ}$. Larger values of $\theta_{12}$ are favoured for  near maximal  $\theta_{23}$. It is worth noting  this dependence (large $\theta_{12}$ associated with near maximal values of $\theta_{23}$) is similar to the lower octant predictions LO 1-3.  For  $49^{\circ}\leq\theta_{23}\leq52^{\circ}$, the $\theta_{12}$ predictions of UO 4 are indistinguishable from  UO 1-3. In spite of this, for certain  $\theta_{23}$, these cases can be differentiated. For example,  $\theta_{23}>52.6^{\circ}$ UO 4 predicts large valued $\theta_{12}$ ($\sim 36^{\circ}$) whereas the $\theta_{12}$ of  UO 1-3 can attain any value in the 3$\sigma$ range. Moreover, at near maximal values of $\theta_{23}$ ($46^{\circ}$), UO 4 predicts smaller $\theta_{12}$ values ($31.3^{\circ}$) than UO 1-3.  Discrimination between UO 1-3 and UO 5 is not possible using $\theta_{12}\mbox{-}\theta_{23}$ correlations (as there is complete overlap in the predictions) and therefore a combination of $\theta_{23}$ and $\delta$ measurements in conjunction with $\nu0\beta\beta$ decay study would be required to disentangle these predictions. \\
 In summary, $\delta$ and $\alpha_{21}$ are bounded between $\pm90^{\circ}$. Moreover, the ability to discriminate between predictions is often dependent upon the value of $\theta_{23}$ and in certain cases predictions are only differentiable with knowledge of the Majorana phases.

\subsection{Predictions Spanning Both Octants}\label{sec:BOP}

We have chosen five  representative cases that span both the upper and lower octants of $\theta_{23}$. The predicted regions of $\theta_{23}$ vary amongst these cases: BO 1 has the greatest viable range, which fully covers the  3$\sigma$ region of $\theta_{23}$. BO 2 and BO 5 also have a wide range of $\theta_{23}$:  $38.2^{\circ}\leq\theta_{23}\leq49^{\circ}$ and $38.2^{\circ}\leq\theta_{23}\leq51^{\circ}$  respectively. BO 3  and BO 4 have the smallest viable range of $\theta_{23}$ with $44.3^{\circ}\leq\theta_{23}\leq53.3^{\circ}$ and $38.2^{\circ}\leq\theta_{23}\leq45.9^{\circ}$ respectively. \\
There is  little structure in the $\delta\mbox{-}\theta_{23}$ correlation of BO 1:  $\delta$ can attain  any value in the range $\pm 90^{\circ}$  and there is no dependence on $\theta_{23}$. BO 2 has a similar  correlation structure   to the lower octant predictions: the maximal $\delta$ value ($73^{\circ}$) is correlated to smaller $\theta_{23}$ values and  CP conserving $\delta$ spans  $42^{\circ}\leq\theta_{23}\leq48.2^{\circ}$. BO 3 and BO 4 have comparable $\delta\mbox{-}\theta_{23}$ dependence;  the  maximal  $\delta$, $69^{\circ}$ and $61^{\circ}$ respectively, occurs at the extreme upper and lower 3$\sigma$ limit of $\theta_{23}$. In comparison with  BO 1-4, BO 5 has a  highly constrained $\delta$ with a maximal value of $14^{\circ}$ for $49^{\circ}\leq\theta_{23}\leq50^{\circ}$. In the scenario $\delta$ is maximally CP violating, the only viable prediction of this set is BO 1. Interestingly, in spite of BO 1 lacking predictivity in regards to  parameters $\delta$, $\alpha_{31}$ and  $\theta_{13}$, its $\alpha_{21}$ and $\theta_{12}$ predictions  attain very specific values ($\alpha_{21}=0^{\circ}$ and $34.8^{\circ}\leq\theta_{12}\leq 35.2^{\circ}$). BO 1  would be of particular interest in $\nu0\beta\beta$ decay studies as it has a single $\alpha_{21}$ value and exceptionally narrow $\theta_{12}$ range. \\
The $\theta_{12}\mbox{-}\theta_{23}$ dependence of BO 3 and BO 4 are  similar: for near maximal  $\theta_{23}$, there is a very limited range of $\theta_{12}$ values ($\sim36^{\circ}$) and for $\theta_{23}$ close to the upper or lower 3$\sigma$ boundary, the possible $\theta_{12}$  become less constrained. This appears to be a common theme of many of the predictions: near maximal $\theta_{23}$ have very specific $\theta_{12}$ predictions.  In the case of BO 2, $\theta_{12}$ can attain any value in the 3$\sigma$ range for $\theta_{23}\leq43.5^{\circ}$ and for larger $\theta_{23}$, smaller values of $\theta_{12}$ are preferable. There is significant overlap in $\theta_{12}$ predictions for BO 2 and BO 5 and only in the scenario, $\theta_{23}\geq 47^{\circ}$ do their predictions differ.  A special feature of BO 5, akin to BO 1, is that certain observables are more constrained than others; for instance predictions of  $\theta_{13}$ and $\theta_{12}$ range widely whereas the  leptonic phases more highly constrained ($|\delta|\leq14^{\circ}$, $|\alpha_{21}|\leq40^{\circ}$, $|\alpha_{31}|\leq90^{\circ}$).\\
In summary,  $\delta$ and $\alpha_{21}$ can only attain values $\pm 90^{\circ}$. Furthermore, there are several examples (BO 1 and BO 5), in which certain observables are  highly unconstrained but in balance other parameters  can only attain very specific values. Therefore in spite of a lack of predictivity in certain observables, these cases still remain testable by upcoming long and medium base-line experiments.

\section{Conclusions}\label{sec:conclusion}

In this article, we studied  the correlations of leptonic observables that result from a flavour symmetry, $A_5$, combined with gCP breaking into residual symmetries $G_{\nu}=\mathbb{Z}_2\times \text{CP}$ and $G_{e}=\mathbb{Z}_2$. This combination of residual symmetries introduces three continuous parameters and unsurprisingly, we obtain a wider range of predictions than in studies that use only one input parameter. The flavour symmetry studies  that implement one input parameter and are of low order such as $A_4$\cite{Feruglio:2012cw,Ding:2013bpa}, $S_4$\cite{Feruglio:2012cw, Feruglio:2013hia} and $A_5$\cite{Li:2015jxa,DiIura:2015kfa,Ballett:2015wia} share common predictions such as  $|\sin \delta|=1$, $|\sin \alpha_{21}|=|\sin \alpha_{31}|=0$  and  maximally CP violating $\delta$ associated with maximal $\theta_{23}$.  We find  the addition of two continuous parameters allows for more possibilities in  correlations  and  predictions of non-trivial leptonic phases differing from $0$, $\frac{\pi}{2}$, $\pi$ and $\frac{3\pi}{2}$. Using a number of example cases we have shown that certain predictions are indistinguishable using oscillation parameters $\delta$, $\theta_{12}$ and $\theta_{23}$ alone  and therefore input from $\nu0\beta\beta$ decay experiments is necessary. We find that, in general, the ability to discriminate between predictions is improved for near maximal $\theta_{23}$ and that even in specific cases in which there is no predictivity for one parameter (e.g. BO 1 and BO 5), other leptonic observables may be highly constrained and provide testable predictions. In spite of a greater number of predictions, all of our cases share the feature of $\delta$ and $\alpha_{21}$ phases being bounded by $\pm 90^{\circ}$, the former of the two which is testable by  long base-line oscillation experiments.\\
In conclusion, we find that relaxing the possible combinations of low-energy residual symmetries permits a wider range of predictions with more complex correlations between leptonic observables  which have the potential to be tested at upcoming neutrino oscillation and $\nu0\beta\beta$ experiments.

\begin{acknowledgements}
We would like to thank  Peter Ballett,  Silvia Pascoli and Serguey Petcov for  advice and invaluable encouragement   throughout this work. In addition, we are grateful for sharing enlightening discussions with Celine Boehm, Mark Ross-Lonergan, Andreas Trautner and Ryan Wilkinson. We would also like to thank Malte Buschmann and  Nick Jennings  for reading various drafts of this paper. This work has been supported by Science and Technology Facilities Council (STFC) and the European Union FP7
ITN-INVISIBLES (Marie Curie Actions, PITN-GA-2011-289442).
\end{acknowledgements}

\begin{figure*}[h!]\label{fig:BOP}
\centering
  \includegraphics[width=1.0\textwidth]{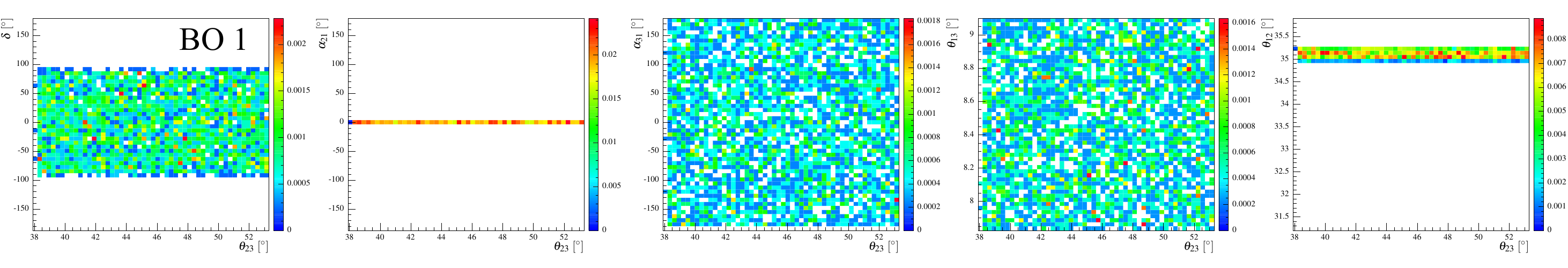}
  \includegraphics[width=1.0\textwidth]{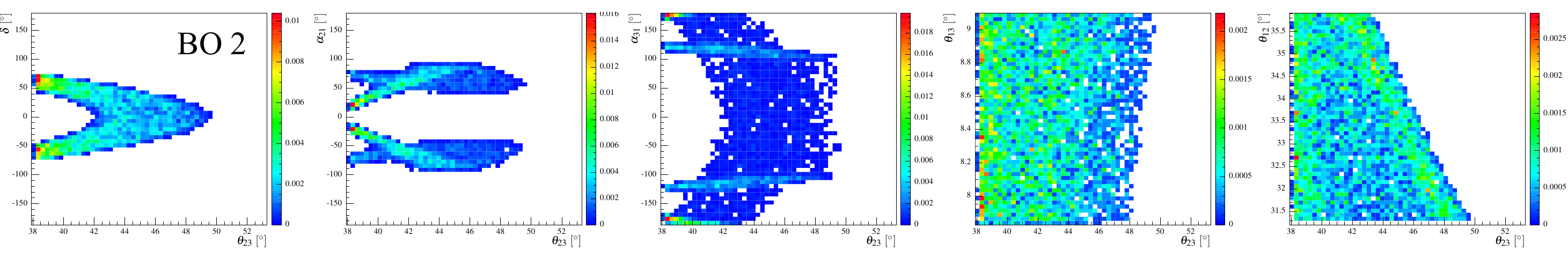}
  \includegraphics[width=1.0\textwidth]{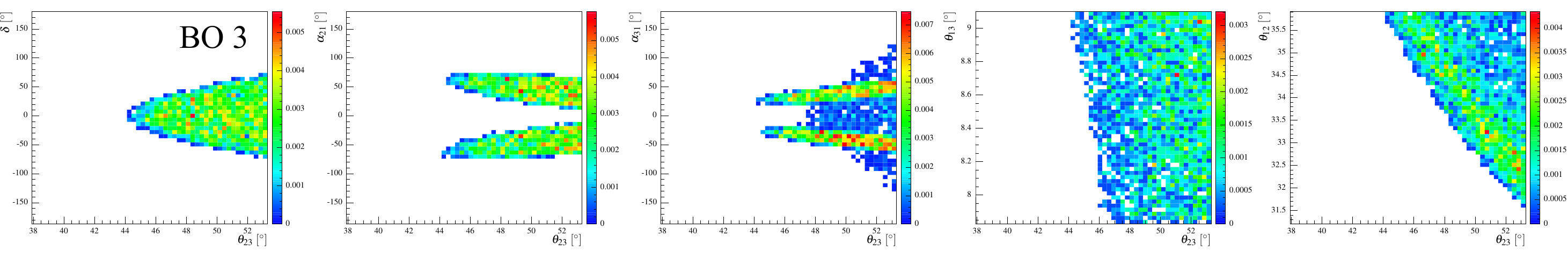}
  \includegraphics[width=1.0\textwidth]{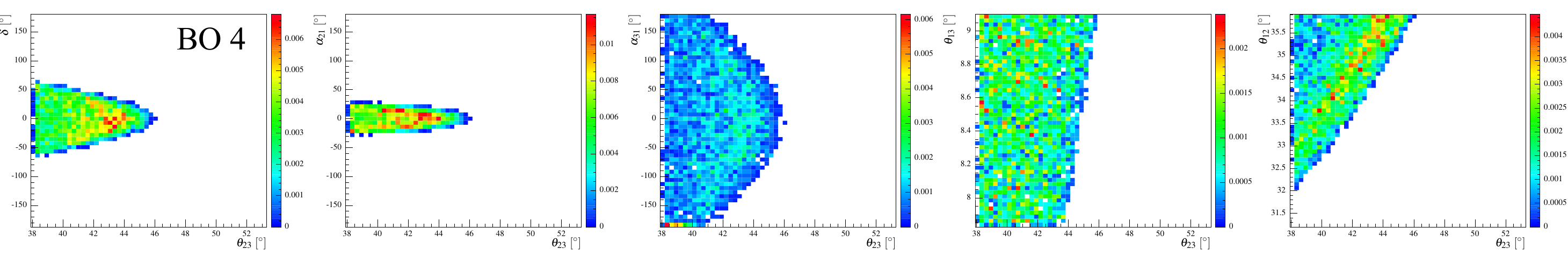}
  \includegraphics[width=1.0\textwidth]{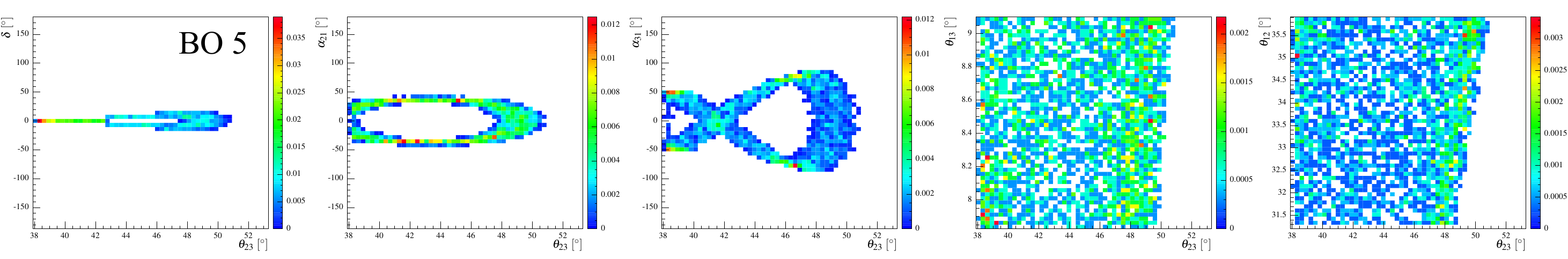}
\caption{Two-dimensional histograms showing the phases  and mixing angles as a function of $\theta_{23}$ for  predictions  agree to a 3$\sigma$ level with global fit data\cite{Gonzalez-Garcia:2014bfa}  The colour chart shows relative frequencies of solutions where red (dark blue) represents a higher (lower)  frequency. Each prediction is labelled 'both octant'  (BO) 1-5.}
\end{figure*}

\section{Appendix}\label{sec:appendix}

In this Appendix we will provide the form of the matrices that we used to derive predictions in the lower, upper and both octant results. The notation that will be used to denote real rotations is,
\begin{equation}
\begin{aligned}
R_{12}&=\begin{pmatrix}
c_{\theta}&s_{\theta}&0\\
-s_{\theta}&c_{\theta}&0\\
0&0&1
\end{pmatrix},\quad
 &R_{13}&=\begin{pmatrix}
c_{\theta}&0&s_{\theta}\\ 
0&1&0\\
-s_{\theta}&0&c_{\theta}
\end{pmatrix}\\\quad\text{and}\quad
R_{23}&=\begin{pmatrix}
1&0&0\\
0&c_{\theta}&s_{\theta}\\
0&-s_{\theta}&c_{\theta}
\end{pmatrix}.
\end{aligned}
\end{equation}

In regards to the complex rotations they will be written as,
\begin{equation}
\begin{aligned}
R_{12}C&=\begin{pmatrix}
c_{\omega}&s_{\omega}e^{i\gamma}&0\\
-s_{\omega}e^{-i\gamma}&c_{\omega}&0\\
0&0&1
\end{pmatrix},\\
R_{13}C&=\begin{pmatrix}
c_{\omega}&0&s_{\omega}e^{i\gamma}\\
0&1&0\\
-s_{\omega}e^{-i\gamma}&0&c_{\omega}
\end{pmatrix}\\\quad\text{and}\quad
R_{23}C&=\begin{pmatrix}
1&0&0\\
0&c_{\omega}&s_{\omega}e^{i\gamma}\\
0&-s_{\omega}e^{-i\gamma}&c_{\omega}
\end{pmatrix}.\\
\end{aligned}
\end{equation}

The $\mathbb{Z}_2$ elements that give distinct results are 
\begin{equation}
\begin{aligned}
\mathbb{Z}_21&=\frac{1}{2}\begin{pmatrix}
\phi&-1&\phi \\
-1&-\phi&-\frac{1}{\phi}\\
\phi&-\frac{1}{\phi}&-1
\end{pmatrix}, \quad
\mathbb{Z}_22&=\frac{1}{2}\begin{pmatrix}
-\phi&-\frac{1}{\phi}&-1\\
-\frac{1}{\phi}&-1&\phi\\
-1&\phi&\frac{1}{\phi}
\end{pmatrix},  \\
\mathbb{Z}_23&=\frac{1}{2}\begin{pmatrix}
-1&-\phi&-\frac{1}{\phi}\\
-\phi&\frac{1}{\phi}&1\\
-\frac{1}{\phi}&1&-\phi
\end{pmatrix}\quad
\mathbb{Z}_24&=\frac{1}{2}\begin{pmatrix}
\frac{1}{\phi}&-1&-\phi\\
-1&-\phi&\frac{1}{\phi}\\
-\phi&\frac{1}{\phi}&-1
\end{pmatrix},\\
\mathbb{Z}_25&=\frac{1}{2}\begin{pmatrix}
-\phi&-\frac{1}{\phi}&1\\
-\frac{1}{\phi}&-1&-\phi\\
1&-\phi&\frac{1}{\phi}
\end{pmatrix},\quad
\mathbb{Z}_26&=\frac{1}{2}\begin{pmatrix}
-1&\phi&\frac{1}{\phi}\\
\phi&\frac{1}{\phi}&-1\\
\frac{1}{\phi}&1&-\phi
\end{pmatrix},\\
\mathbb{Z}_27&=\frac{1}{2}\begin{pmatrix}
-1&\phi&\frac{-1}{\phi}\\
\phi&\frac{1}{\phi}&-1\\
\frac{-1}{\phi}&-1&-\phi
\end{pmatrix}.\\
\quad
\end{aligned}
\end{equation}
\newline
The diagonalising matrix of $\mathbb{Z}_2\text{i}$ will be denoted by $U_{\text{i}}$ for \text{i} $\in 1..6$.
The permutations that have been applied to account for the arbitrariness of ordering of the eigenvectors will be denoted by $\text{p}_1..\text{p}_6$,

\begin{equation}
\begin{aligned}
p_1&=\begin{pmatrix}
1&0&0\\
0&1&0\\
0&0&1
\end{pmatrix},\quad 
p_2&=\begin{pmatrix}
1&0&0\\
0&0&1\\
0&1&0
\end{pmatrix},\\
p_3&=\begin{pmatrix}
0&1&0\\
1&0&0\\
0&0&1
\end{pmatrix},\quad
p_4&=\begin{pmatrix}
0&1&0\\
0&0&1\\
1&0&0
\end{pmatrix}\\
p_5&=\begin{pmatrix}
0&0&1\\
1&0&0\\
0&1&0
\end{pmatrix},\quad
p_6&=\begin{pmatrix}
0&0&1\\
0&1&0\\
1&0&0
\end{pmatrix}\\
\end{aligned}
\end{equation}

\subsection{Lower Octant Matrices}
\begin{table}[H]
\centering
\begin{tabular}{ c | c  }
 Result  & $U_{\text{PMNS}}$ \\
\hline\hline

LO 1 & $p_{4} R_{23}C {U_{1}}^{\dagger}\Omega_{12}R_{12}p_2$\\
LO 2& $ p_{2}R_{13}C {U_{2}}^{\dagger}\Omega_{12}R_{13} p_{3}$\\
LO 3 & $p_{1} R_{12}C {U_{3}}^{\dagger}\Omega_{12}R_{23}p_4$\\
LO 4& $ p_{2}R_{13}C {U_{2}}^{\dagger}\Omega_{12}R_{13} p_{1}$\\
LO 5& $p_4 R_{23}C {U_{1}}^{\dagger}\Omega_{12}R_{13}p_3$\\

\end{tabular}
\end{table}

\subsection{Upper Octant Matrices}

\begin{table}[H]
\centering
\begin{tabular}{ c | c  }
 Result  & $U_{\text{PMNS}}$ \\
\hline\hline
UO 1 & $p_{1} R_{13}C {U_{2}}^{\dagger}\Omega_{12}R_{13}p_1$\\
UO 2& $ p_{1}R_{13}C {U_{4}}^{\dagger}\Omega_{12}R_{13} p_{3}$\\
UO 3 & $p_{4} R_{12}C {U_{3}}^{\dagger}\Omega_{12}R_{23}p_4$\\
UO 4& $ p_{3}R_{23}C {U_{5}}^{\dagger}\Omega_{12}R_{13} p_{3}$\\
UO 5& $p_3 R_{23}C {U_{1}}^{\dagger}\Omega_{12}R_{12}p_2$\\
\end{tabular}
\end{table}

\subsection{Both Octant Matrices}

\begin{table}[H]
\centering
\begin{tabular}{ c | c  }
 Result  & $U_{\text{PMNS}}$ \\
\hline\hline
BO 1 & $p_{1} R_{23}C {U_{5}}^{\dagger}\Omega_{12}R_{12}p_4$\\
BO 2 & $ p_{1}R_{12}C {U_{6}}^{\dagger}\Omega_{13}R_{12} p_{4}$\\
BO 3 & $p_{4} R_{23}C {U_{1}}^{\dagger}\Omega_{23}R_{12}p_6$\\
BO 4 & $ p_{4}R_{12}C {U_{7}}^{\dagger}\Omega_{23}R_{23} p_{1}$\\
BO 5 & $p_{3}R_{23}C {U_{5}}^{\dagger}\Omega_{13}R_{13}p_3$\\
\end{tabular}
\end{table}

\clearpage
\bibliographystyle{apsrev4-1}
\bibliography{main}{}

\end{document}